\documentclass[12pt,preprint]{aastex}  

\slugcomment{Accepted for publication in ApJ}
\shorttitle{Local Interstellar Magnetic Field}\shortauthors{Frisch}
\def\glong{$\ell$}
\def\glat{$b$}
\def\elong{$\lambda$}
\def\elat{$\beta$}
\def\PAgal{PA$_\mathrm{G}$}

\def\PAecl{PA$_\mathrm{E}$}
\def\PAoneecl{PA$_\mathrm{S1,E}$}

\def\NH{$N$(H)}
\def\NHI{$N$(H$^\circ$)}
\def\NaI{Na$^\circ$}
\def\NCaII{$N$(Ca$^{\rm +}$)}

\def\NHII{$N$(H$^{\rm +}$)}
\def\nHI{$n$(H$^{\circ}$)}
\def\nHII{$n$(H$^{+}$)}
\def\CII{C$^{\rm +}$}

\def\HI{H$^{\rm o}$}
\def\DI{D$^{\rm o}$}
\def\FeII{Fe$^{\rm +}$}
\def\NFeII{$N$(Fe$^{\rm +}$)}
\def\CaII{Ca$^{\rm +}$}
\def\CaIII{Ca$^{\rm ++}$}
\def\MgII{Mg$^{\rm +}$}
\def\MgI{Mg$^{\circ}$}
\def\OI{O$^{\circ}$}
\def\HII{H$^{\rm +}$}

\def\HeI{He$^{\rm o}$}
\def\nel{$n_\mathrm{e}$}
\def\P5{$P_5$}
\def\Btheta{$\mathrm{B}_\theta$}
\def\Bphi{$\mathrm{B}_\phi$}

\def\kms{\hbox{km s$^{-1}$}}
\def\deeg{\hbox{$^{\rm o}$}}
\def\lya{\hbox{Ly$\alpha$}}
\def\cmtwo{cm$^{-2}$}
\def\cc{cm$^{-3}$}

\begin{document}                                                                

\title{The S1 Shell and Interstellar Magnetic Field and Gas near the Heliosphere} 
\author{Priscilla C. Frisch}
 \affil{Department of
Astronomy and Astrophysics, University of Chicago, Chicago, IL 60637.}

\begin{abstract}

Many studies of the Loop I magnetic superbubble place the Sun at the
edges of the bubble. One recent study models the polarized radio
continuum of Loop I as two magnetic shells with the Sun embedded in
the rim of the 'S1' shell.  If the Sun is in such a shell, it should
be apparent in both the local interstellar magnetic field and the
distribution of nearby interstellar material.  The properties of these
subshells are compared to the interstellar magnetic field (ISMF) and
the distribution of interstellar \FeII\ and \CaII\ within $\sim 55$ pc
of the Sun.  Although the results are not conclusive, the ISMF
direction obtained from polarized stars within $\sim 30 $ pc is
consistent with the ISMF direction of the S1 shell.  The distribution
of nearby interstellar \FeII\ with log \NFeII$<12.5$ \cmtwo\ is
described equally well by a uniform distribution or an origin in
spherical shell-like features.  Higher column densities of \FeII\ (log
\NFeII$>12.5$ \cmtwo) tend to be better described by the pathlength of
the sightline through the S1 and S2 subshells.  Column densities of
the recombinant ion \CaII\ are found to increase with the strength of
the interstellar radiation field, rather than with star distance or
total pathlength through the two magnetic subshells.  The ion \CaII\
can not be used to trace the distribution of local interstellar gas
unless the spatial variations in the radiation field are included in the
calculation of the ionization balance, in addition to possible
abundance variations.  The result is that a model of Loop I as
composed of two spherical magnetic subshells remains a viable
description of the distribution of nearby low density ISM, but is not
yet proven.

\end{abstract}

\section{Introduction} \label{sec:intro}

The location of the Sun in the rim of the Loop I superbubble has been
inferred from radio continuum data, kinematical data on the flow of
local ISM away from the center of Loop I, data on gas-phase abundances
in local ISM, and the coincidence of the velocity of ISM inside and
outside of the heliosphere.  Loop I is an evolved superbubble shell
formed from stellar evolution in a subgroup of the Sco-Cen
association, $\sim 4-5$ Myrs ago
\citep[e.g.][]{deGeus:1992,Frisch:1995,Frisch:1996,MaizApel:2001}.
Both the original dimensions found for the Loop I bubble observed in
820 MHz \citep{Berkhuijsen:1973}, and more recent studies of
\citet[][H98a,H98b]{Heiles:1998lb,Heiles:1998whence} and
\citet{Wolleben:2007}, place the Sun in or adjacent to the rim of a
magnetic superbubble shell for an assumed spherical geometry.  The 1.4
GHz Wolleben study defines two magnetic subshells of Loop I, S1 and
S2, with magnetic pole directions differing by $90^\circ \pm
42^\circ$.  Comparisons between the radio continuum filaments of Loop
I and optical polarization data indicate that the radio filaments at
distances of $\sim 100 - 150$ pc trace magnetic field lines,
indicating that optical polarization is a suitable tracer of magnetic
shells (H98a).  Both the kinematics and abundance pattern of local
interstellar material (LISM) suggest that the Loop I remnant has
expanded to the solar location \citep{Frisch:1981}.  LISM abundances
of the refractory elements Mg, Fe, and Ca, show the characteristic
enhancement indicative of grain destruction in interstellar shocks
\citep{Frischetal:1999}.  Local interstellar gas, $d < 50$ pc, and
dust flow away from the center of Loop I at a best-fit velocity of
$\sim 18 $ \kms\ in the local standard of rest \citep[LSR,
e.g.][]{Frischetal:2009ibex}.  The first spectrum of backscattered
Ly$\alpha$ emission from interstellar hydrogen inside of the
heliosphere showed that the velocity of interstellar \HI\ inside of
the heliosphere is comparable to LISM velocities
\citep{AdamsFrisch:1977}.  Together these data suggest that the
magnetic field and spatial configuration of the LISM can be used to
test whether the Loop I magnetic superbubble has expanded to the solar
location.  The \citet[][]{Wolleben:2007} model of the S1 and S2 shells
provides enough detail to make preliminary comparisons between LISM
data and the properties of these shells.  These comparisons provide
interesting insights into the LISM properties, and support the
possibility that local ISM within $\sim 55$ pc is dominated by the S1
and S2 shells.

Superbubble expansion into ambient ISM with equal magnetic and thermal
pressures yields roughly spherical superbubbles during early
expansions stages when magnetic pressure is weak compared to the ram
pressure of the expanding gas
\citep{MacLowMcCray:1988,FerriereZweibel:1991}, and bubbles elongated
along the ISMF during late stages of evolution
\citep{HanayamaTomiska:2006snrsb}.  The evolved shell is thicker near
the ISMF equatorial regions, where field strengths are larger due to
flux freezing, than the polar regions of the shell where thermal
pressure provides the main support for the shell.  In media where
magnetic pressure is weak, e.g. the ratio of thermal to magnetic
pressure $\beta > 10$, the evolved bubble is more symmetric.
Supernovae in Sco-Cen Association subgroups have contributed to the
evolution of the Loop I superbubble during the past $\leq 14$ Myrs.
The Loop I superbubble (and S1, S2) expanded in a medium with a
density gradient, because the initial supernova occurred in the
molecular regions of the parent Scorpius-Centaurus Association
subgroups, while the subsequent bubble expansion occurred in the low
density interior of the Local Bubble cavity
\citep{Frisch:1981,Frisch:1995,Fuchsetal:2006}. In this case the
external plasma $\beta$ may have varied irregularly across the
expanding shell, so that the topology of the present day S1 and S2
shells may deviate from axial symmetry as well as sphericity.

The ISMF direction at the heliosphere provides the most direct measure
of whether the Sun is embedded in the shell of the Loop I superbubble.
Several phenomena trace the field direction -- the weak polarization
of light from nearby stars \citep[][hereafter
F07]{Tinbergen:1982,Frisch:2007cmb}, the flield direction in the S1
subshell of Loop I \citep[]{Wolleben:2007}, the 3 kHz emissions from
the outer heliosheath detected by the two Voyager satellites
\citep[][F07]{GurnettKurthetal:2006}, the observed angular offset
between interstellar \HI\ and \HeI\ flowing into the heliosphere
\citep{Lallementetal:2005,PogorelovZank:2006,Opheretal:2007}, and the
10 pc difference between the distances of the solar wind termination
shock detected by the two Voyager satellites
\citep[e.g.][]{Stone:2008}.  The orientation of the plane midway
between the hot and cold dipole moments of the cosmic microwave
background is also within $\sim 15^\circ$ of the local ISMF direction
(F07).
\footnote{More recently, IBEX has found that the ISMF interacting with
the heliosphere forms a ribbon of energetic neutral atom emission as
viewed from the Earth, and the ribbon traces regions where the ISMF
direction within the outer heliosheath is perpendicular to the sightline
\citep[e.g.][]{McComasetal:2009science}.}

This paper searches for evidence that the S1 and S2 shells affect the
distribution of nearby ISM within $\sim 55$ pc.  The topology of the
S1 and S2 shells is
discussed in \S \ref{sec:shells}.  Section \ref{sec:mag} shows that
the direction of the ISMF at the Sun is consistent with the ISMF
direction in the S1 shell, similar to the location of the mid-plane
between the cosmic microwave dipole moments, and consistent with the ISMF
direction inferred from heliosphere models.  The distribution of the
ISM in the S1 and S2 shells are compared to \FeII\ column densities
towards nearby stars behind the shells (\S \ref{sec:feII}).  A similar
comparison is made between the \CaII\ data and the S1 and S2 shells,
however \CaII\ column densities appear instead to trace the strength
of the local far ultraviolet (UV) diffuse radiation field (\S
\ref{sec:caII}).  An appendix outlines the ionization equilibrium
of \CaII.

\section{Approximating the Three-Dimensional ISM Distribution in the S1 and S2 Shells } \label{sec:shells}

\citet[][]{Wolleben:2007} has fit two separate spherical magnetic
shells ('S1' and 'S2') to the low frequency (1.4 GHz and 23 GHz)
polarized radio continuum, which must have a relatively local origin
because of the $\lambda^2$ dependence of Faraday rotation.  The ISMF
is assumed to be entrained in the expanding superbubble shell, with no
deviation from spherical symmetry.  The Sun is located in the rim of
the S1 shell, which is centered $ 78\pm 10$ pc away at galactic
coordinates $\ell,b= 346^\circ \pm5^\circ , 3^\circ \pm 5^\circ $.
The upwind direction of the flow of local ISM past the Sun is within
$\sim 20^\circ$ of of the S1 shell center.\footnote{Comparisons
between the LSR flow velocity and Loop I require using a somewhat
uncertain velocity correction to obtain the LSR motion of the cloud.
For the 'Standard' LSR, the local ISM flows at a velocity of --19.4
\kms\ from the direction of \glong,\glat=331\deeg,--5\deeg, while an
LSR correction based on Hipparcos star distances gives a bulk flow
velocity of --17 \kms, from \glong,\glat=2\deeg,--5\deeg\
\citep{FrischSlavin:2006astra}.}  The inner and outer radii of the S1
shell are $72 \pm 10$ and $ 91 \pm 10$ pc respectively.  Wolleben
described the S1 magnetic field direction by two angles, the angle
between the field direction and the NGP \Bphi $= -72^\circ \pm
30^\circ$, and the rotation about the NGP \Btheta $=71^\circ \pm
30^\circ$.  The S2 shell center is more distant ($\sim 95 $ pc) and
centered at higher galactic latitudes ($b \sim 37^\circ$) than the S1
shell, with an ISMF direction near the north galactic pole.  An
alternate single-shell model for Loop I is also based on the \HI\
shell and centers the feature at \glong,\glat$=320^\circ,5^\circ$
(H98a).  As a first approximation of shell structure, they are assumed
spherically symmetric, although the ISMF is seen in filaments
interacting with denser clouds for more distant regions of Loop I
(H98a) A more detailed model of these evolved bubbles requires
understanding the magnetic pressure.  The parameters provided for the
S1 and S2 shells by Wolleben are detailed enough for comparison with
observations of the LISM.

\begin{figure}[h!]
\plottwo{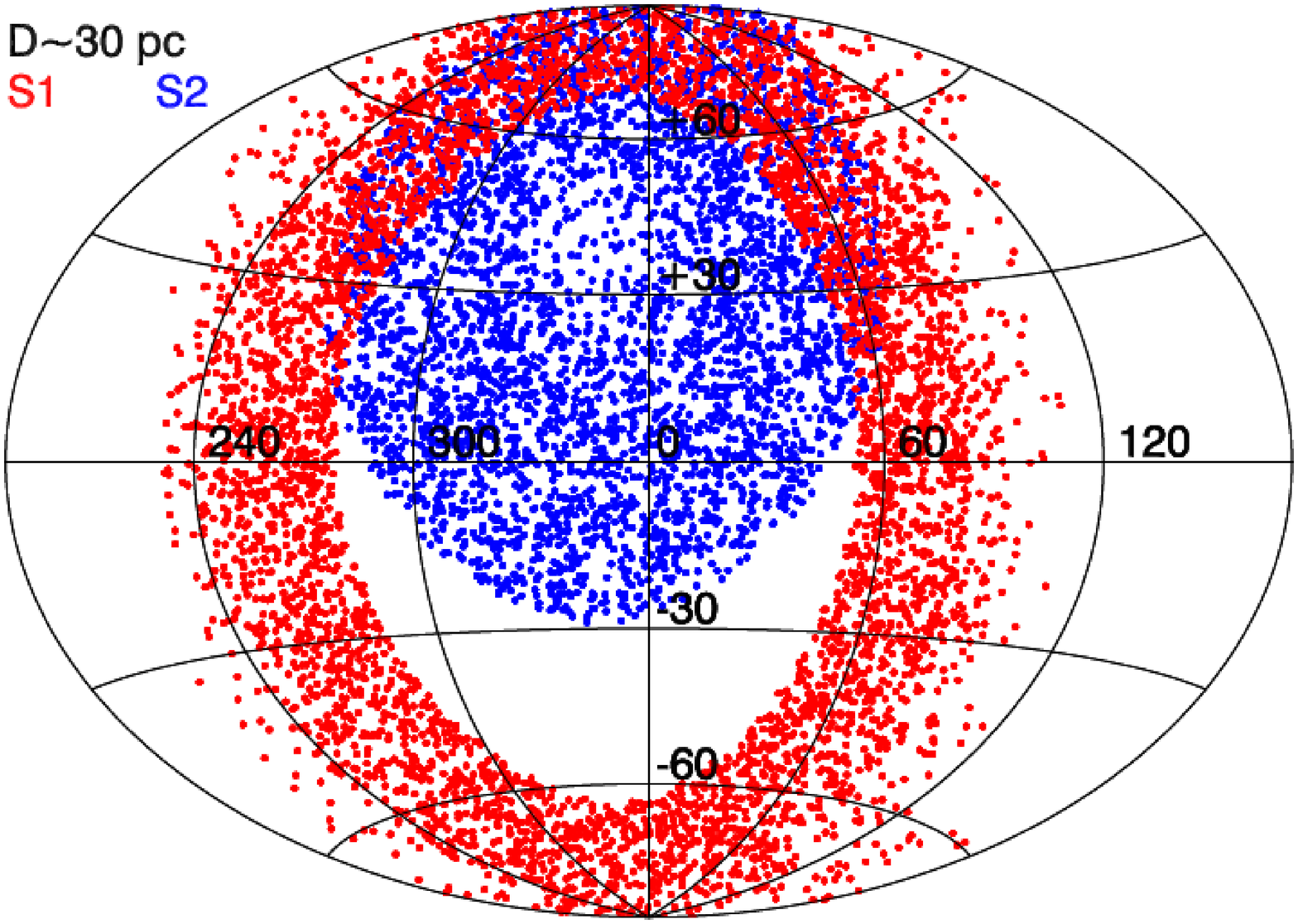}{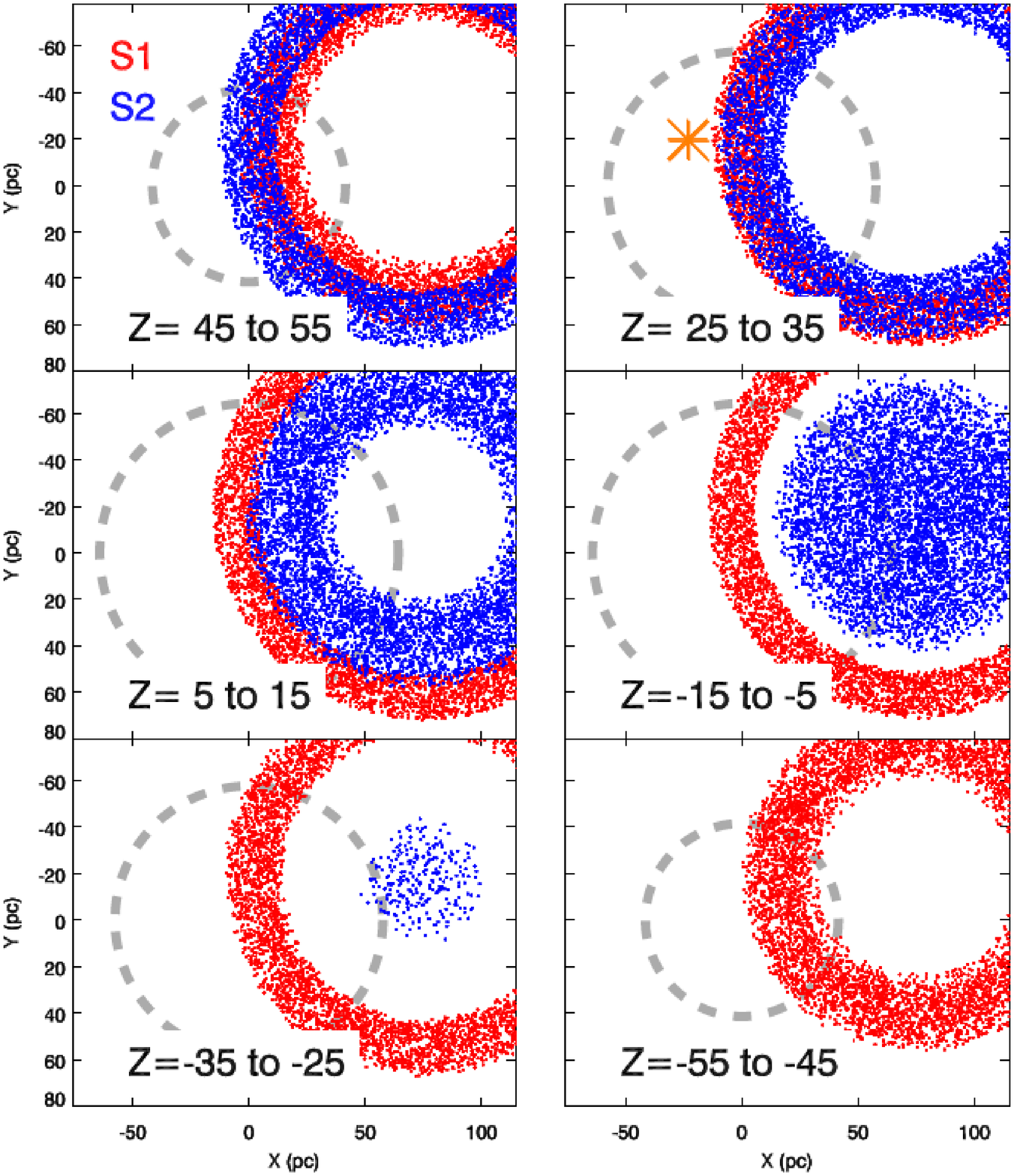}
\caption{ Left: The parts of the S1 and S2 shells about 30 pc from the
Sun.  Right: Distribution of points in the S1 (red) and S2 (blue)
shells for different distance intervals Z (pc) above and below the
galactic plane.  The galactic center is at y=0 and x=infinity, beyond
the plot right.  The direction of galactic rotation is at x=0,
y=infinity, beyond the plot bottom.  The individual points trace the
range of ISM distributions in the shells calculated with the
assumption that each shell is spherically symmetric, and that the ISM
is uniformly distributed.  The dotted gray lines show the distance
corresponding to 55 pc from the Sun for the midpoint of the
Z-interval.
The S1 shell distribution is for a shell centered at
\glong,\glat$=346^\circ, 3^\circ$ and 78 pc, with a radius of 82 pc,
and with a total rim thickness of 19 pc.
The S2 shell distribution is for a shell centered at
\glong,\glat$=347^\circ, 37^\circ$ and a distance of 95 pc, with a
radius of 75 pc and total rim thickness of 24 pc.  The orange star
shows the location of the nearby cold filament towards Leo
\citep{Meyer:2007,Lauroesch:2007}. 
\label{fig:slices} }
\end{figure}

A three-dimensional (3D) spherically symmetric model of the S1 and S2
shells is created for comparison with the local interstellar magnetic
field and distribution of interstellar \FeII\ and \CaII.  The 3D
configuration is initially constructed in the frame of each shell such
that the north pole is at the local zenith. The shell rims are filled
with a uniform density of points.  The shell model is then rotated to
the galactic coordinate system by \Btheta\ and \Bphi, and translated
to the shell center in the galactic coordinate system.  The result is
a model where a path through the shell measures the shell column
density normalized to an arbitrary value, and the shell column density
in any sightline
varies according to the end point of the path.  Fig. \ref{fig:slices},
right, shows the ISM distribution in the S1 and S2 shells for slices
parallel to the galactic plane and for 10 pc-wide intervals of Z above
and below the galactic plane.  Fig. \ref{fig:slices}, left, shows
sections of the two shells at distances of $\sim 30$ pc using an
aitoff projection.  Negative Z-values are dominated by the S1 shell
(red), below Z$\sim 30$ pc (latitudes $b \sim 20 ^\circ$). At higher
latitudes a given sightline may sample either, or both, of the shells.
The parameters of the 3D simulation place the Sun in the rim of the S1
shell, but the uncertainties quoted by Wolleben also allow the Sun to
be in the S2 shell.  As additional data on nearby ISM become
available, it should become possible to both constrain and test the
S1/S2 models in more detail.

The S1 and S2 shells overlap in places.  One example is the sightline
towards the stars $\alpha$ Oph (HD 159561), located 14.3 pc away at
\glong,\glat$=35.9^\circ,22.6^\circ$, which has the strongest \CaII\
line observed towards any nearby star \citep{Crawford:2001}.  The S1
and S2 shells coincide at a distance of 12 pc in this sightline,
suggesting that the \CaII\ line towards $\alpha$ Oph samples a region
where the S2 shell collided with the S1 shell, possibly creating a
shock so that recent grain destruction occurred.  Merging flows induce
thermal instabilities that generate such filamentary structures
\citep{AuditHennebelle:2005}, and the more distant \HI\ gas in this
sightline is also filamentary.  A second example of possible
interacting shells is the nearby Leo filament \citep{Lauroesch:2007}.
The orange symbol in Fig. \ref{fig:slices} (Z=25--35 pc) shows the
location of a tiny cold (20 K) filamentary ($> 7^\circ \times
2^\circ$) cloud in Leo, located at \glong,\glat$=220^\circ,45^\circ$
and at a distance of less than 42 pc
\citep{Meyer:2007,Lauroesch:2007}.  If the cloud is at 40 pc, it is
outside of both shells for the basic values for the shell distances
and radii (i.e. without invoking any uncertainties).  However if the
cloud is nearby, or if the extreme values allowed by the uncertainties
on the S1 and S2 shells are invoked, this filament may form where the
two shells collide.  Merging flows induce thermal instabilities that
generate such filamentary structures as the Leo filament
\citep{Lauroesch:2007,AuditHennebelle:2005}, so the presence of this
cold filament is consistent with the picture of the LISM as dominated
by two shell features.

\section{The S1 and S2 Shells and Local Interstellar Magnetic Field } \label{sec:mag}

\subsection{S1 Shell and ISMF Direction}\label{sec:tinb}

\begin{figure}[h!]
\plottwo{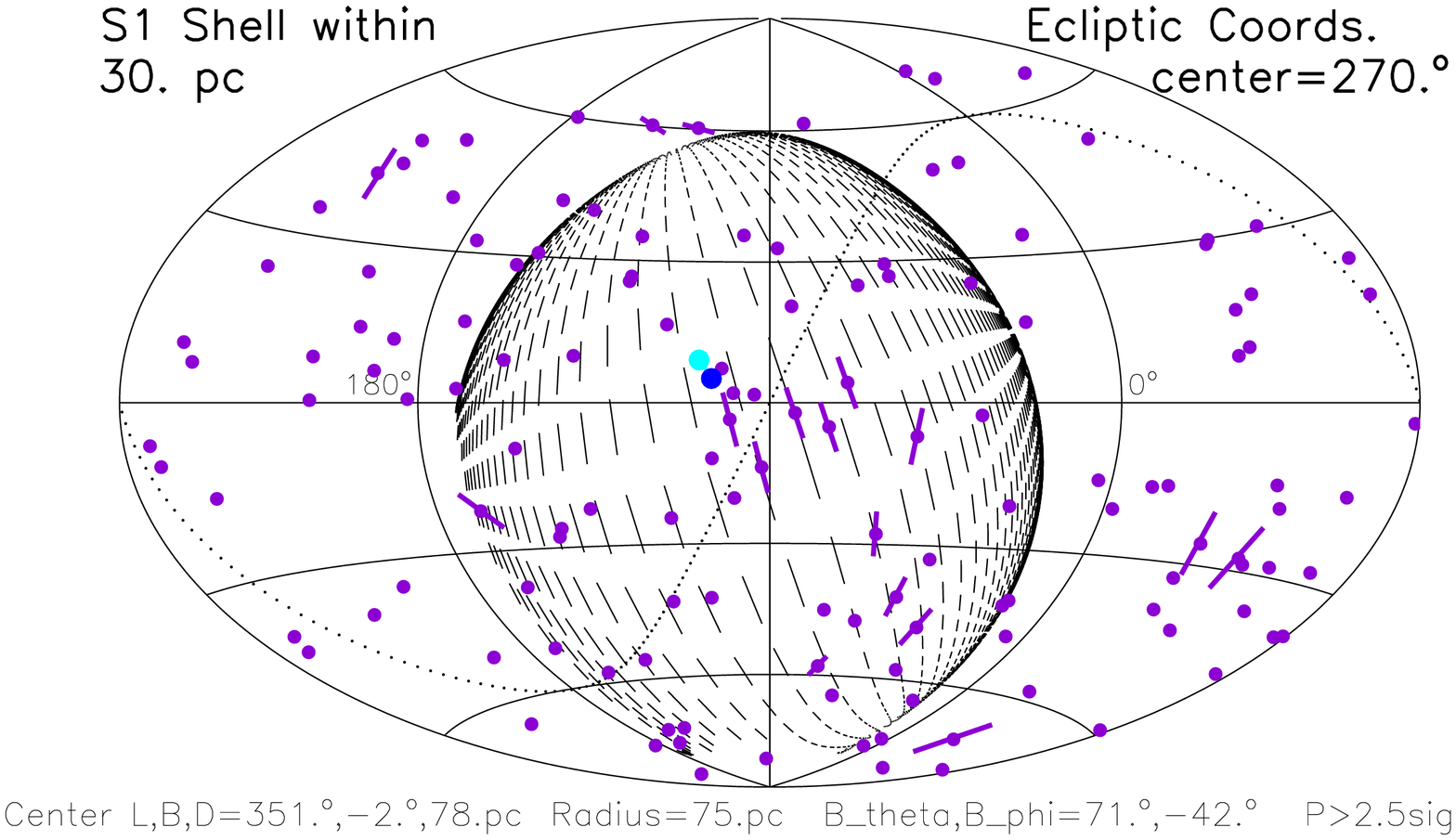}{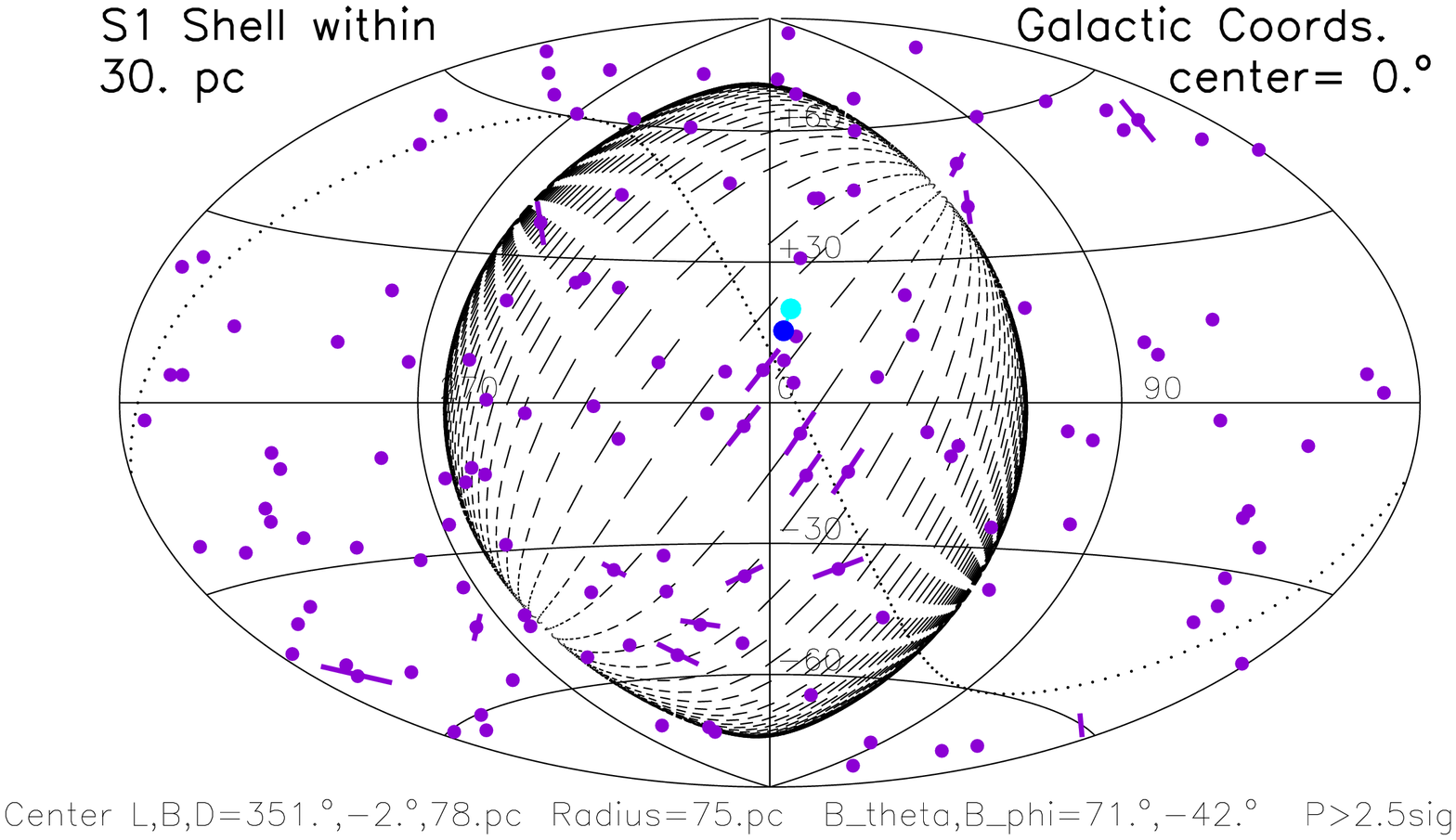}
\caption{ The magnetic field associated with the parts of the S1 shell
within 30 pc is plotted in ecliptic coordinates (left) and galactic
coordinates (right) for an aitoff projection.  The ecliptic plot is
centered at $\lambda = 270^\circ$, while the galactic plot is centered
at \glong=0\deeg.  The parameters for the S1 shell given in
\citet{Wolleben:2007} have been varied within the range of allowed
uncertainties to yield the best match to the \citet{Tinbergen:1982}
polarization data towards stars in the heliosphere nose region.  The
dark and light blue dots show the inflow directions of interstellar
\HI\ and \HeI\ into the heliosphere.  The S1 subshell parameters used
in the above figures correspond to a shell center at
(\glong,\glat)=$(351^\circ,-2^\circ)$ and 78 pc, shell radius of 75
pc, and magnetic field angles \Btheta=71\deeg, \Bphi=--42\deeg.  The
dots show stars within 50 pc with polarization data, and the red bars
show polarization vectors for stars where \P5$> 2.5 \sigma$
\citep[][]{Tinbergen:1982,Piirola:1977,Frisch:2007cmb}.
The longitude increases towards the right for both figures.
\label{fig:tinb} }
\end{figure}

Polarization by charged irregularly shaped interstellar grains yields
polarization vectors that are parallel to the ISMF, because the
induced magnetic torques naturally align the grains
\citep[e.g.][]{Lazarian:2000}.  There is a patch of dust towards the
fourth galactic quadrant (\glong$>270^\circ$), mainly in the southern
hemisphere and in the upwind direction of the interstellar gas flowing
through the heliosphere, where the ISMF within 5--40 pc of the Sun has
been traced by very weak optical polarizations \citep{Tinbergen:1982}.
The Tinbergen data were acquired in the southern hemisphere during
1974, and northern hemisphere data during 1973 (J. Tinbergen, private
communication).  Tinbergen detected polarizations of $\gtrsim
0.017$\%, with $1 \sigma \sim 0.007 $\%, towards a few stars within 40
pc.  Five of these stars
\footnote{The five stars with strongest polarizations in the
heliosphere nose region are HD 161892, HD 177716, HD 181577, HD
155885, and HD 169916, with the first three stars showing polarization
detections at the $3 \sigma$ level.}  are close to the ecliptic plane,
and offset by up to \elong$\le 40 ^\circ$ from the heliosphere nose
towards positive ecliptic longitudes
\citep{Frisch:2005L,Frisch:2007cmb}.  The mean position angles for the
three stars near the nose with $P> 3\sigma$, are \PAgal=$33^\circ \pm
11^\circ$ in galactic coordinates, and \PAecl=$ - 26^\circ \pm 11
^\circ$ in ecliptic coordinates (Table 1).  The position angle
uncertainty is estimated by allowing Q and U to vary over $\pm 1
\sigma$ (Table \ref{tab:stars}).  The nearest star towards the nose is
36 Oph, at 6 pc, and it has a $2.5 \sigma$ detection with $P=0.018$\%.
The polarization position angle of 36 Oph (\PAecl=--19.9, \PAgal=39.5)
does not differ significantly from the mean position angle of the
three more distant stars with $3\sigma$ polarizations in the nose
region.  The polarizations of the Tinbergen's sample, together with
northern hemisphere data of \citet{Piirola:1977}, are plotted in Fig.
\ref{fig:tinb}.  An alternative catalog of nearby star polarizations
is the comprehensive catalog assembled by \citet{Leroy:1993a}; however
it is mainly based on measurements with larger uncertainties, and is
therefore less useful for identifying very low polarization levels.

In order to test the relation between the ISMF in the S1 shell and the
Tinbergen polarization data, the ISMF direction for the S1 shell was
varied within the uncertainties on \Btheta\ and \Bphi\ to find the
direction that is the most consistent with the optical data.  The best
match to the optical polarization data was found for
\Btheta$=71^\circ$ and \Bphi$=-42^\circ$.  The parts of the S1 shell
within 30 pc are compared to the optical polarization data in
Fig. \ref{fig:tinb}, for galactic (right) and ecliptic (left)
coordinates, for stars within 50 pc; polarization vectors are plotted
if $P> 2.5 \sigma$.  For the purpose of this figure, I use a center
position for the S1 shell of (\glong,\glat)$=(351^\circ,-2^\circ)$ and
radius 75 pc, so that the Sun is located 3 pc outside of the shell
rim.  The mean position angle of the S1 magnetic field (for
\Bphi=--42\deeg) at the star locations is \PAoneecl$=-20 \pm 5 $ in
ecliptic coordinates, which is within the uncertainties of the stellar
polarizations.  A slightly larger value of \Bphi\ exactly matches the
mean position angles of the starlight polarizations, but violates the
quoted uncertainties.  Therefore, the ISMF directions from the
Tinbergen data and S1 shell configuration are consistent to within the
uncertainties.

The Tinbergen stars with the strongest polarizations are $\sim
90^\circ$ from the pole of the S1 shell, which is consistent with the
expectation of higher ISMF field strengths where the shell expansion
is perpendicular to the ISMF direction.  The best value for the S1
magnetic field direction close to the Sun, derived from comparisons
with these optical polarization data (\S \ref{sec:mag}), is
\Btheta$=71^\circ$ and \Bphi$=-42^\circ$, corresponding to a local
ISMF direction towards \glong,\glat$=71^\circ,48^\circ$.

Heliospheric asymmetries are caused by interactions with
the interstellar magnetic field.
A widely used measure of the heliosphere distortion due to the ISMF,
which is inclined by the angle $\alpha \sim 30^\circ-60^\circ$ with
respect to the ISM flow vector, is the observed offset between the
inflowing \HeI\ and \HI\ directions
\citep{Witte:2004,Lallementetal:2005}.  The ISMF also shifts the
maximum \lya\ emission originating in the outer heliosheath
\citep{BenJaffelRatkiewicz:2000}.  Correcting the \HeI\ and \HI\
directions to a common observation epoch yields an offset angle separation of $4.9^\circ \pm
1^\circ$ between the two directions.  (The upwind direction
of the \HeI\ flow in J2000 coordinates is \elong$=255.4^\circ \pm
0.5^\circ$, \elat$= 5.1^\circ \pm 0.2^\circ$, Witte, private
communication).  These directions define a
position angle, which is \PAecl$= -35^\circ \pm 20^\circ$
(\PAgal$=26^\circ \pm 20^\circ$) in ecliptic (galactic) coordinates,
respectively.  Large uncertainties are quoted because the upwind
direction of the \HI\ flow through the heliosphere is not precisely
defined due to the $\sim 50$\% of the interstellar \HI\ lost to
filtration in the hydrogen wall, the balance between radiation
pressure and gravity affecting trajectories of \HI\ atoms surviving to
the heliosphere interior, and the production of secondary \HI\ atoms
inside of the heliosphere \citep[e.g.,][]{QuemeraisIzmodenov:2002}.
For comparison, at the heliosphere nose location,
\glong,\glat$=3.5^\circ,15.2^\circ$, the S1 shell with
\Bphi=--42\deeg\ gives a position angle \PAecl=$-15^\circ$.  The
position angle formed by the offset between \HeI\ and \HI\ flowing
through the heliosphere is marginally consistent with the S1 shell
direction at the heliosphere nose.  The 10 AU difference in the
termination shock distance found by Voyagers 1 and 2, in 2004 and 2007
respectively \citep{Stone:2008}, must be combined with the \HI-\HeI\
offset to provide a more reliable constraint on models of the
direction of the interstellar magnetic field affecting the heliosphere
\citep[e.g.][]{PogorelovStoneetal:2007,Opheretal:2007}.
\footnote{Since this paper was originally submitted, a number of
recent papers have appeared that discuss the ISMF direction at the
Sun, based on MHD heliosphere models
\citep[e.g.][]{Ratkiewiczetal:2008,Pogorelovetal:2008let}, IBEX data
on the ENA Ribbon \citep{Schwadron:2009sci,Funsten:2009sci}, or both
\citep{Heerikhuisen:2010ribbon}.  There is an overlap between ISMF
directions in these models and the uncertainties on Wolleben's ISMF
direction for the S1 shell.  The overlap occurs for galactic
longitudes in the range of $40^\circ - 50^\circ$ and galactic
latitudes in the range of $23^\circ - 42^\circ$.  The models also
predict an ISMF direction that is directed towards negative ecliptic
latitudes.}

\subsection{S1 shell and the CMB Dipole Moment}

The great circle that is midway between the hot and cold poles of the
cosmic microwave background (CMB) dipole passes within $5^\circ$ of
the interstellar \HeI\ upwind direction, and bifurcates the
heliosphere nose \citep{Frisch:2007cmb}.  The point of closest
approach to the nose is at \glong,\glat$=7.4^\circ,11.6^\circ$, where
the position angle of the CMB dipole mid-plane in ecliptic coordinates
is \PAecl$= -11^\circ \pm 1^\circ$ (uncertainties in the upwind
\HeI\ direction, used to define the nose position,
are included in this uncertainty, Table 1).  For comparison, the S1
shell magnetic field direction at this location, for \Bphi$=-42^\circ$, corresponds to a
position angle (ecliptic coordinates) of \PAecl=$-18^\circ$.  This
fact is mentioned here because the low-$\ell$ multipole moments of the
CMB show symmetries related to the ecliptic geometry
\citep[e.g.][]{Copietal:2006}, so that the symmetry of the CMB dipole
moment around the heliosphere nose should also be of interest.  The
upwind direction is $5^\circ$ above the ecliptic plane, so any
coincidence between the CMB multipole moments and the ecliptic
geometry is tantamount to a coincidence with the heliosphere
morphology and/or to the interstellar magnetic field that shapes the
heliosphere.  These ecliptic signatures on the CMB are not understood,
but it is not unreasonable to postulate they arise from processes
related to the local interstellar magnetic field and its effect on the
heliosphere.
\footnote{In results published after the submission of this paper, it
has been shown that the ISMF interacting with the heliosphere controls
the flow of nanometer-sized interstellar dust grains around and
through the heliosphere \citep{SlavinFrisch:2009sw12}. }


\section{Comparisons between S1 and S2 Shells and Distribution of \FeII\ } \label{sec:feII}

If the distribution of nearby ISM is determined by the S1 and S2
magnetic superbubbles, then the S1 and S2 shell morphologies should be
imprinted on the strengths of interstellar absorption lines.
Fig. \ref{fig:slices} shows two views of the ISM associated with the
S1 and S2 shells.  Interstellar absorption lines towards stars within
55 pc show that most of the the LISM is warm, $\sim 3,000-12,000$ K
\citep{RLIII} with low average spatial densities, $<0.1$ \cc.
Exceptions are the tiny dense clouds occasionally seen in \NaI\ and
\HI\ absorption \citep{Meyer:2007}.  Models of the radiative transfer
properties of the circumheliospheric ISM show a partially ionized, low
density cloud, \nHI$\sim 0.20$ \cc, \nel$\sim 0.07$ \cc, and with
temperature $\sim 6,300$ K determined from interstellar \HeI\ inside
of the heliosphere \citep[][SF08]{Witte:2004,SlavinFrisch:2008}.
Therefore, a suitable tracer of the S1 and S2 shell morphologies
should be abundant, insensitive to cloud ionization, and undepleted.
There are no available data sets that meet all three requirements,
therefore the criteria that the element be undepleted is dropped.  The
best element for this study is then \FeII, which has been measured
towards $\sim 27$ stars within 56 pc
\citep{Lehneretal:2003,RLI,Kruketal:2002}.  Iron is predominantly
singly ionized in the cloud around the heliosphere, with neutral and
Fe$^{++}$ together containing less than 3\% of the Fe atoms (SF08).  A
more difficult aspect of using \FeII\ to trace absolute ISM densities
is the factor of $\sim 6-40 $ difference in the gas-phase \FeII\
abundances between dense cold and warm tenuous clouds, due to dust
grain destruction by interstellar shocks including in local regions
\citep[e.g.][]{SlavinJones:2004}.  The alternative common element that
traces both neutral and ionized gas is \MgII, however it has similar
abundance variations as \FeII\ and the \MgII\ h and k lines may be
more saturated than the \FeII\ lines.  Therefore, I use \FeII\ column
densities to trace the distribution of local ISM.  Three nearby stars
are omitted from this discussion because they have known debris disks
\citep[HD 215789, HD 209952, HD 216956,][]{debrisdisks:2006}.

\begin{figure}[h!]
\plottwo{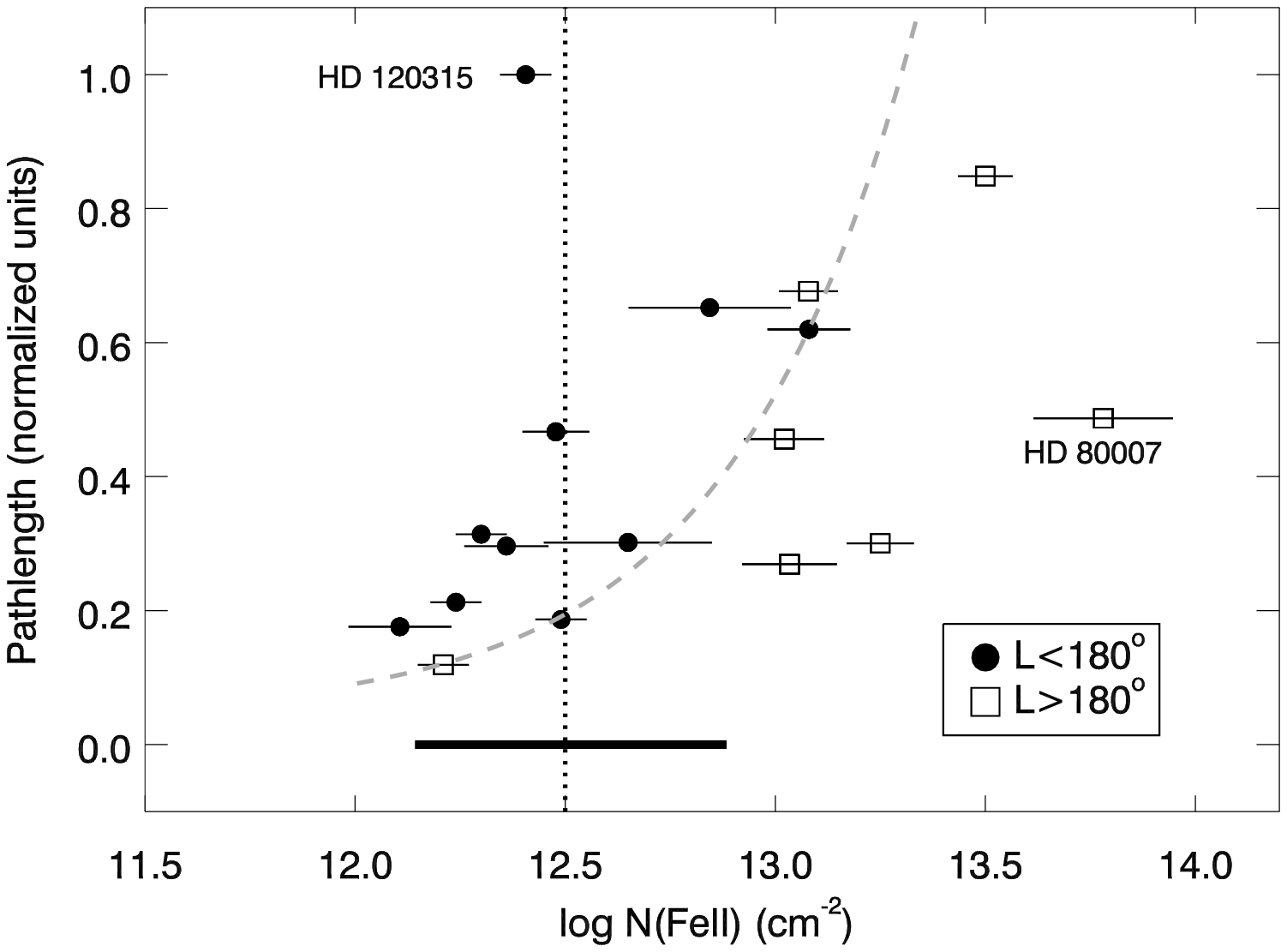}{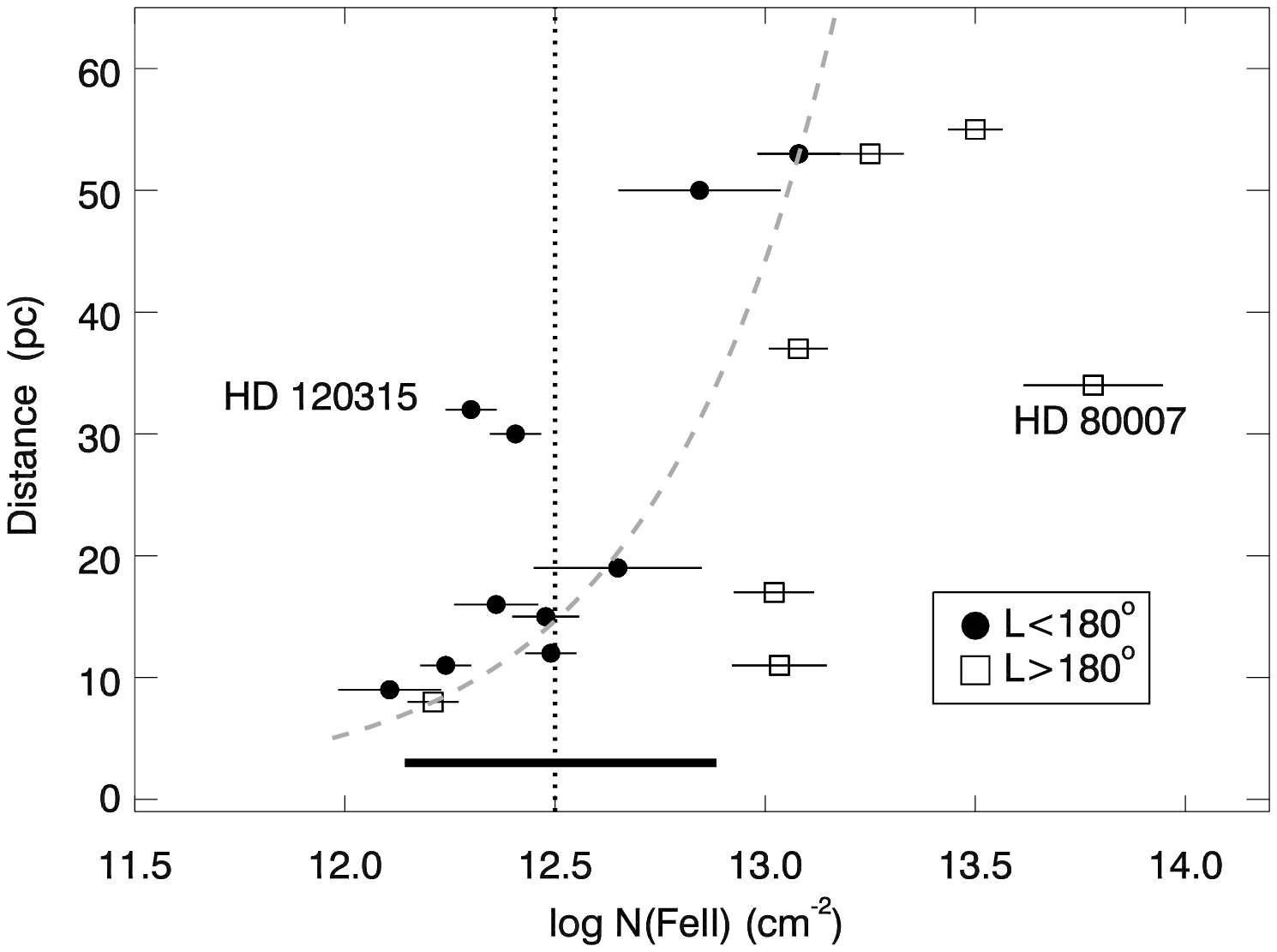}
\caption{Column densities of \FeII\ are plotted against the total
predicted pathlength through the S1 and S2 shells foreground to the
stars (or equivalently column densities, in arbitrary units) through
the S1 and S2 shells (left, see \S \ref{sec:feII}) and against the
star distance (right).  The symbols indicate stars with stars 6--56 pc
away and with galactic longitude \glong$>180^\circ$ (open squares) and
\glong$<180^\circ$ (filled dots).  The horizontal bar at the bottom shows the
column density range for the stars within 6 pc that are not
plotted.  The vertical
line separates stars with log $N$(\FeII)$<12.5$ \cmtwo\ versus $>12.5$
\cmtwo\, where column densities tend to increase with distance or
pathlength through the shells, respectively.  The stars are listed in
Fig. \ref{fig:panels}.  The dashed line shows a first order fit
between the total column densities and pathlength (left), and total
column densities and star distance (right).  The individual velocity
components for each star are added to give the total \NFeII\ plotted
on the abscissa.  The highest \FeII\ column density in this sample is
towards $\beta$ Car (HD 80007), 34 pc from the Sun in the direction
\glong,\glat$=286^\circ,-14^\circ$.  According to Fig. 1, right,
and Fig. 4, left,
this star samples a tangential pathlength through the S1 shell.}
\label{fig:length} 
\end{figure}

\begin{figure}[th!]
\plottwo{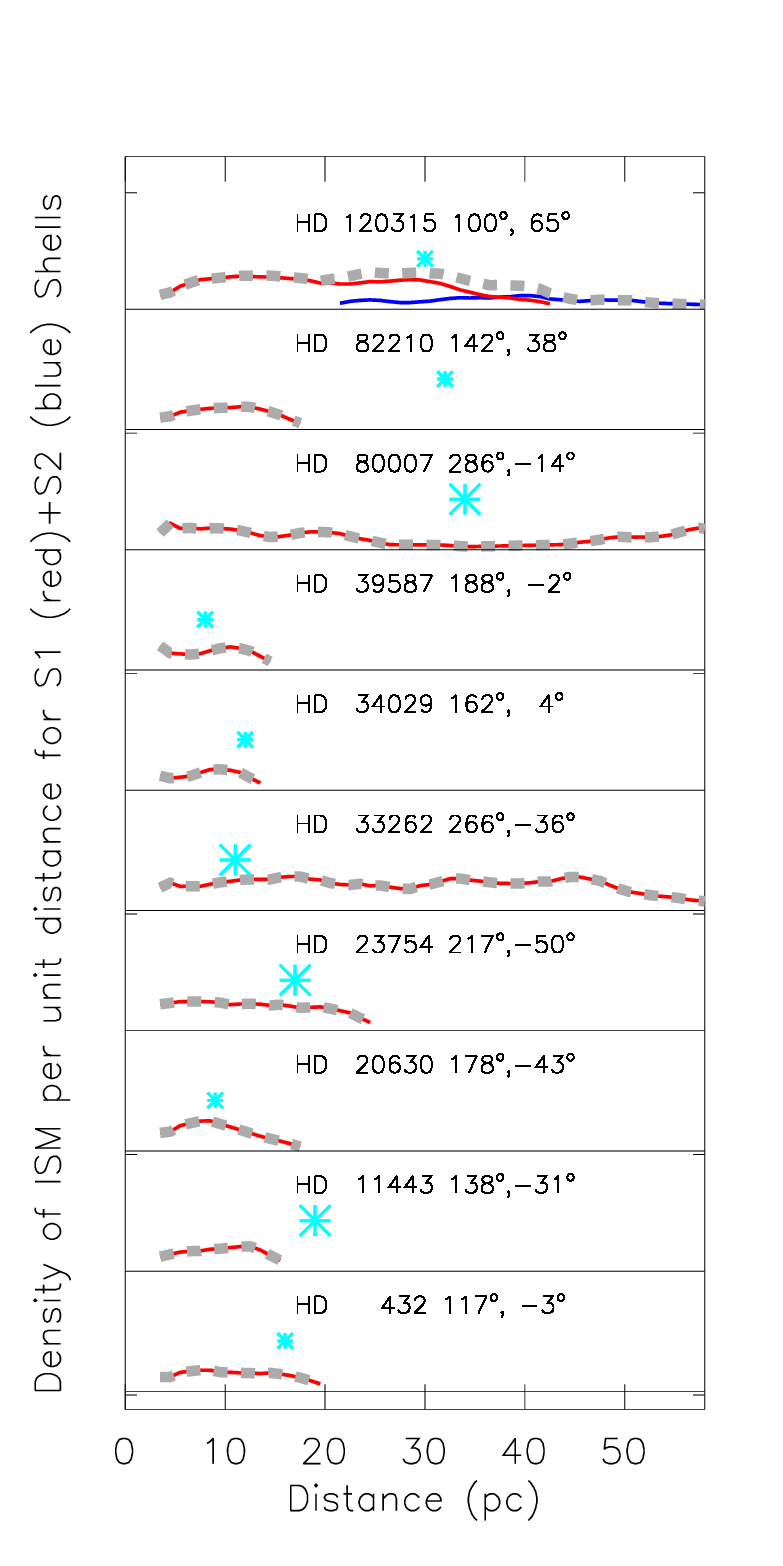}{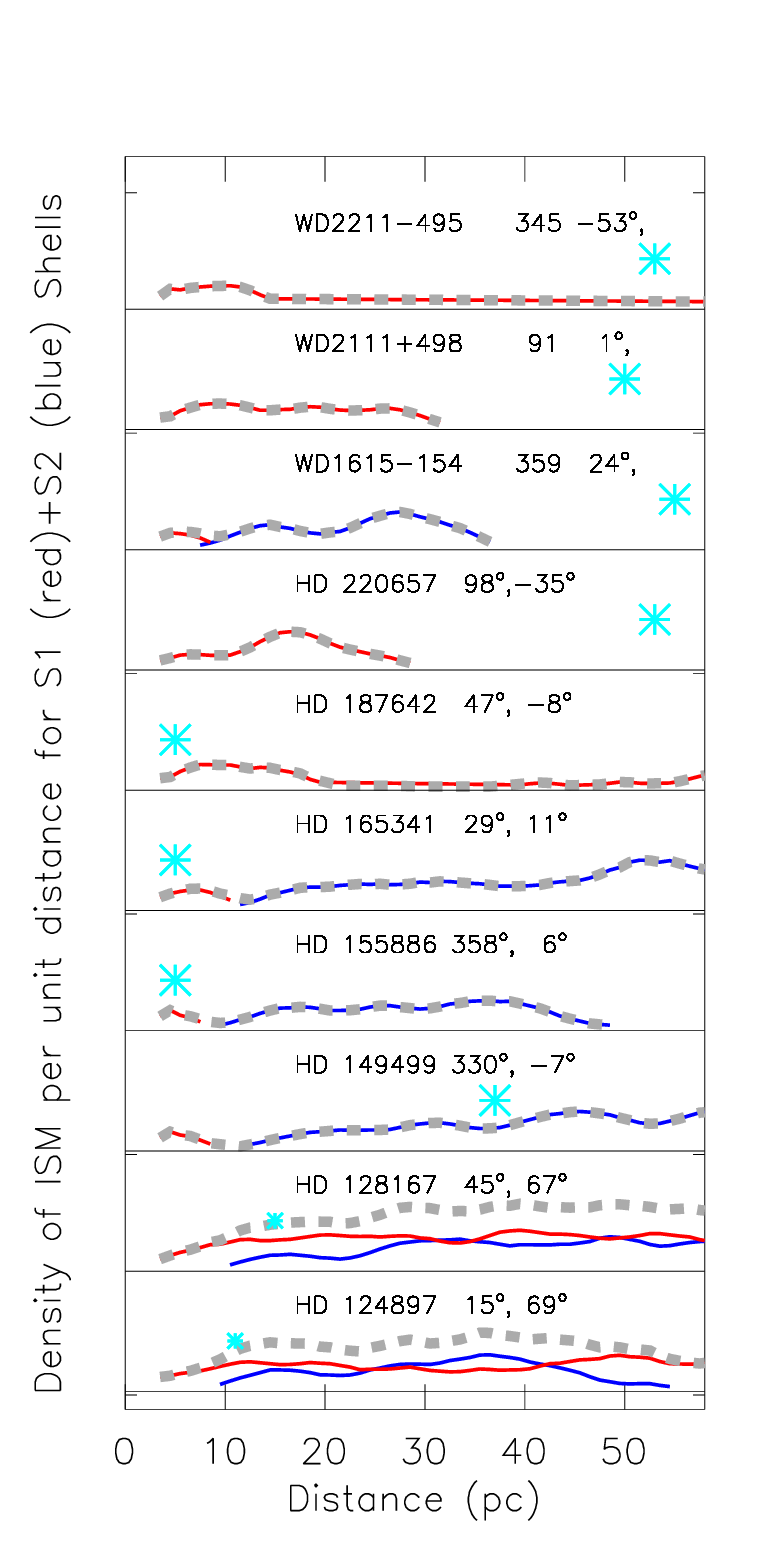}
\caption{ The distributions of ISM in the S1 shell (red) and S2 shell
(blue) are shown towards each star in Fig. \ref{fig:length}, based on
a spherical morphology for each shell (see text). The distributions
are plotted for arbitrary density units.  The sum of the S1 and S2
densities is plotted as the thick gray dotted line.  The turquoise symbol
shows the distance of the star (labeled by name and coordinates), and
small (large) symbol sizes show log \NFeII$< 12.5 $ \cmtwo\ ($> 12.5 $
\cmtwo) towards the star.
\label{fig:panels}}
\end{figure}

Comparisons between log \NFeII\ and the column density of shell gas
(S1 and S2) in front of each star, versus a comparison between \NFeII\
and the star distance, provide useful insights
(Fig. \ref{fig:length}). The shell column density towards each star
represents the sum of the densities through the parts of the S1 and S2
shells foreground to the star, normalized to an arbitrary value.  This
pathlength was constructed by assuming a column width of $\pm 5$ pc in
order to smooth out uncertainties in the intrinsic shell parameters
\citep[Table 1 in ][]{Wolleben:2007}.  Because of this smoothing,
stars within 6 pc are omitted from Fig. \ref{fig:length}.  The bar at
the bottom of the figures shows the column density range for stars
within 6 pc.  For low column densities, log \NFeII $< 12.5$ \cmtwo,
\FeII\ column densities tend to increase with star distance.  For
higher column density sightlines, log \NFeII $> 12.5$ \cmtwo, \NFeII\
clusters more tightly around the shell pathlength (e.g. column
density) than around the star distance.  The exceptions to both
comparisons are the stars HD 120315 (a low column density,
high-latitude star that should sample a long pathlength through the
shells) and HD 80007 (a high column density low-latitude star, with a
path that is tangential to the S1 shell).  The \FeII\ components in a
sightline are summed together for this comparison.  The dashed lines
in Fig. \ref{fig:length} show a linear fit between the column
densities and the ordinate.  The individual stars are listed in
Fig. \ref{fig:panels}, where the relative column densities of the S1
and S2 shells towards each star are shown.

A second property of the distribution of \FeII\ in these figures is
that stars with galactic longitudes $> 180^\circ$ tend to have larger
\FeII\ column densities, at a given distance, than stars in the
opposite hemisphere. This effect is not seen in \DI\ or \HI.  One
possible reason is that the Fe abundances differ between the two
hemispheres.  The problem with this explanation is that it requires
abundance variations over spatial scales of several parsecs, in
relatively low velocity ISM ($<20$ \kms\ LSR), with the variations
ordered by the arbitrary coordinate of galactic longitude.  An
alternative explanation is that the \FeII\ lines for
\glong$<180^\circ$ include cool unresolved clumps of ISM.  The ISRF
towards \glong$>180^\circ$ is significantly larger than towards
\glong$<180^\circ$ (\S \ref{sec:caII}), so that ISM for
\glong$<180^\circ$ experiences reduced heating because of the
absorption of H-ionizing photons, which may allow unresolved ISM
clumps to coexist with warmer gas at the same velocity.  Once the
ionization of local ISM is better understood, the morphology of the S1
and S2 shells can be adjusted to better represent the actual
distribution of the local ISM.

The stars within 6 pc show total column densities log \NFeII$=12.14 -
12.88$ \cmtwo.  Of these stars, the strongest lines are towards HD
155885, HD 165341, and HD 187642 in the upwind direction.

The distributions of S1 and S2 shell material towards each star are
shown in Fig. \ref{fig:panels}.  Stars with high (low) column
densities are plotted as large (small) symbols for each direction.
The most evident property is that the S1 shell dominates the ISM
towards the southerly stars.  These can be used to predict whether a
star (or exoplanet system for example) will be embedded in a
cloud-like feature or in the Local Bubble plasma.  The assumed
spherical morphology for S1 and S2 leads to the predictions that the
white dwarf HD149499B (WD1634-573) is embedded in in the relatively
denser gas of the shell, while the white dwarf WD1615-154 will be
embedded in the low density Local Bubble plasma.

\section{\CaII\ and Electron Densities in the S1, S2 Shells} \label{sec:caII}

\begin{figure}[h!]
\plottwo{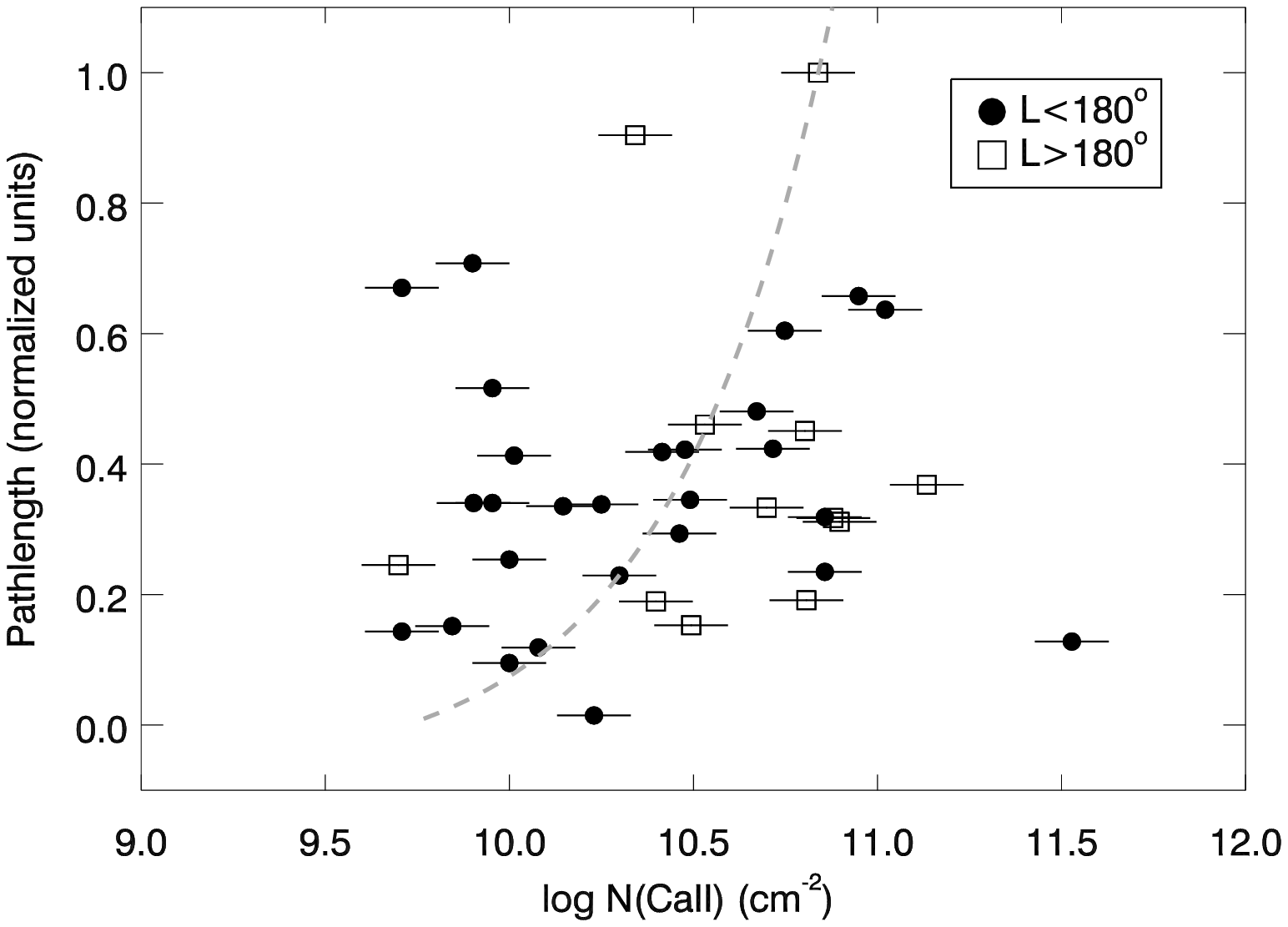}{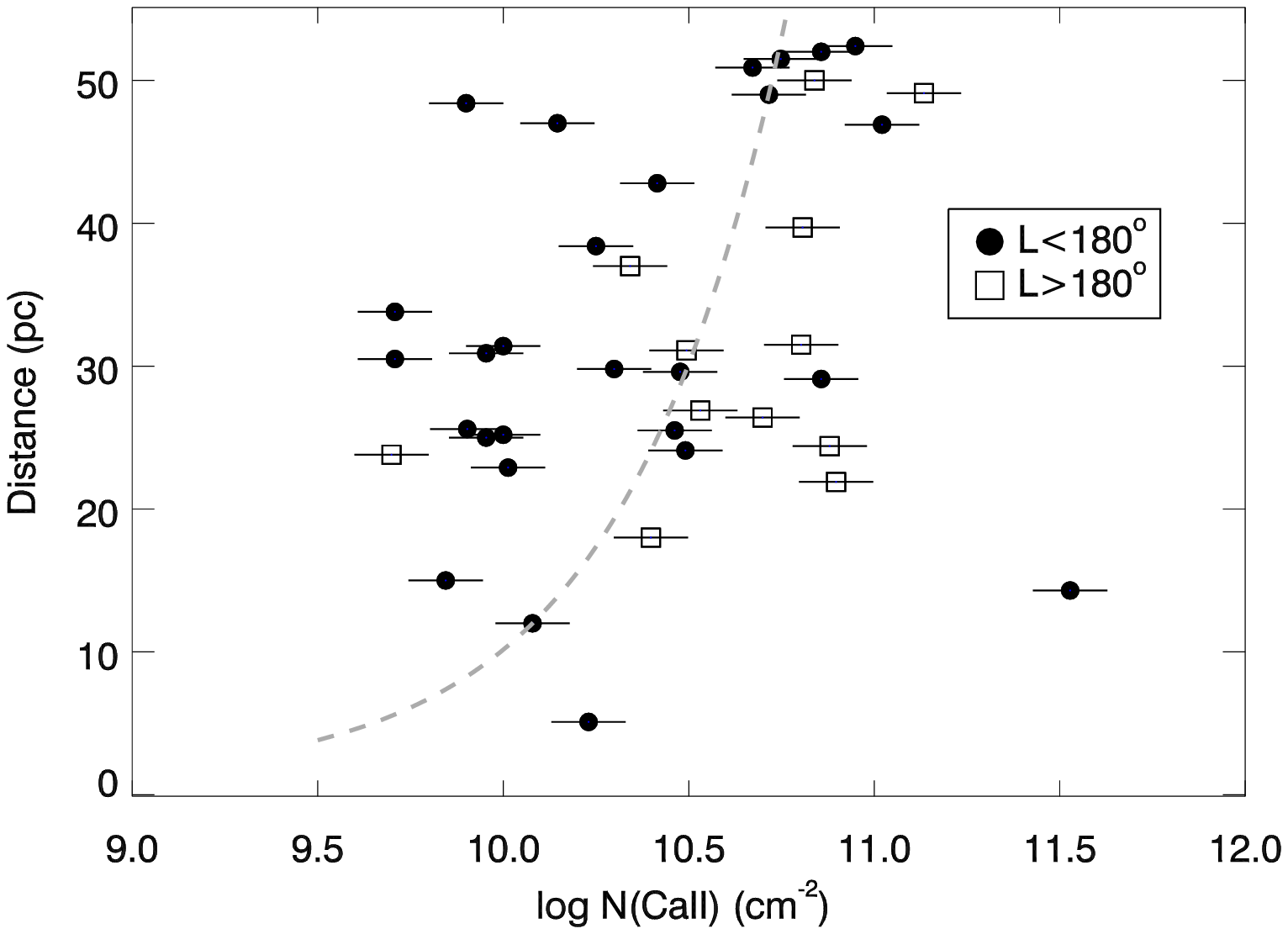}
\caption{Left: Column densities of \CaII\ are plotted against the
total predicted pathlength (in arbitrary units) through the S1 and S2
shells sampled by each star.  The stars are 5--55 pc away and are
coded by whether the galactic longitude is \glong$>180^\circ$ (open
squares) or \glong$<180^\circ$ (filled dots).  The dotted line shows a
first order fit between the total column densities and pathlength
(left), and total column densities and star distance (right).  The
first two galactic quadrants, \glong$=0^\circ - 180^\circ$ show
generally smaller \CaII\ column densities than do stars with
\glong$=180^\circ - 360^\circ$.  The individual velocity components
for each star are added to give the total \NCaII\ plotted on the
absicissa.  Right: The same set of stars are plotted against the
distance of the star from the Sun.  }
\label{fig:lengthcaII} 
\end{figure}

Interstellar \CaII\ is a recombinant species that traces the electron
density as well as abundance variations.  Because \CaII\ is formed
through recombination, it is a proxy for the electron density in
nearby low density gas providing that abundance variations and the
radiation field are understood.  If the ISRF and electron densities
are uniform throughout the S1 and S2 shells, then \CaII\ column
densities would show a similar dependence on pathlength through the
shells as seen for \FeII.  Interstellar \CaII\ column densities are
plotted against the pathlength through the S1 and S2 shells
(Fig. \ref{fig:lengthcaII}, left) and star distance
(Fig. \ref{fig:lengthcaII}, right), using data from
\citep{FrischChoi:2008ip,FGW:2002,WeltyCa:1996}.  The \CaII\ column
densities do not correlate with either the star distance or pathlength
through the S1 and S2 shells.  Sightlines where the \CaII\ column
density is an upper limit are not included in this comparison.

Higher \CaII\ column densities are found for stars with
\glong$>180^\circ$, as was seen for \FeII.  The mean \CaII\ column
density is $\sim 40$\% higher for stars in Quadrants III and IV,
\glong$>180^\circ$, compared to stars with \glong$<180^\circ$. The
difference becomes a factor of two if the anomalously strong \CaII\
line towards $\alpha$ Oph is ignored.  For \glong$<180^\circ$,
$<$\NCaII$>= 3.7 \times 10^{10} $ \cmtwo\ (23 stars).  Omitting
$\alpha$ Oph, which has the strongest known \CaII\ for nearby stars
\citep[e.g.][]{Crawford:2001}, gives a mean for the \glong$<180^\circ$
sample of $<$\NCaII$> = 2.3 \times 10^{10}$.  For \glong$>180^\circ$,
$<$\NCaII$> = 5.1 \times 10^{10} $ \cmtwo\ (23 stars).  No similar
effect is seen in \HI\ (or \DI) column densities towards nearby stars
\citep[based on data in][]{Woodetal:2005}.  For stars within 50 pc,
the mean \HI\ column density does not vary between the \glong$<
180^\circ$ hemisphere and the \glong$> 180^\circ$ hemisphere, and both
show a mean value of \NHI$\sim 1.5 \times 10^{18} $ \cmtwo\ for
respective sample sizes of 25 and 27 stars.

A different picture emerges when \CaII\ (and \FeII) column densities
are compared to the far UV radiation flux at the star.  The highest
diffuse far UV fluxes are seen towards stars in the third and fourth
galactic quadrants, \glong$>180^\circ$, because of the low ISM opacity
in the Local Bubble interior, hot stars in the Scorpius-Centaurus
Association, and $\alpha$ Vir
\citep[][Go80,OW84]{Gondhalekaretal:1980,OpalWeller:1984}.  Fluxes at
975 A are given for the 25 brightest stars, based on a survey by the
$STP~72-1$ satellite in the 910--1050 A band and flux models (OW84).
At least 97\% of the local flux at 975 A is provided by the stars with
\glong$>180^\circ$.  Using the 975 A data for the 25 brightest stars
(OW84), the 975 A flux was calculated at each star and is plotted
against \CaII\ column densities (Fig. \ref{fig:caII}, left), and
\FeII\ column densities (Fig. \ref{fig:caII}, right).  The \FeII\ and
\CaII\ samples are different, with more than half of the \FeII\ stars
within 20 pc, while most of the \CaII\ stars are beyond 20 pc.  All of
the stars in the \FeII\ data set with 975 A flux levels larger than
$6.8 \times 10^4$ photons cm$^{-2}$ s$^{-1}$ A$^{-1}$ are in galactic
Quadrants III and IV, and these stars tend to have larger \FeII\
column densities.  The ISRF gradient would affect \NFeII\ only through
larger column densities of HII, since \FeII\ dominates in both ionized
and neutral diffuse ISM.  Nearby H II gas in regions with
\glong$>180^\circ$ would explain \FeII variations, without violating
the \HI\ constraints (which are determined from a similar star
sample).

The dependence of \CaII\ column densities on the 975 flux is less
simple because \CaII\ is a trace species formed by recombination, with
an ionization potential of 11.87 eV versus 13.60 eV for \HI.  In the
cloud around the Sun, $n \sim 0.2$ \cc\ and \CaIII/\CaII=63 (SF08).
The ISRF gradient near the Sun (Fig. \ref{fig:caII}) affects \NCaII\
several ways.  Higher radiation fluxes lead to higher electron
densities, increasing \CaIII$\rightarrow$\CaII\ recombination, and the
overall \HII\ fraction would increase.  Higher radiation fluxes also
increase the \CaII\ photoionization rate, but this effect does not
appear to be dominant.  In the local ISM, photoionization appears to
dominate over collisional ionization.  Radiative transfer models of
the ISM surrounding the Sun match available ionization data such as
the \MgII/\MgI\ ratio (SF08).  Low observed Ar$^\circ$ abundances
towards nearby stars also indicate the dominance of photoionization
\citep{SofiaJenkins:1998}.  The radiation flux at 1044 A capable of
ionizing \CaII\ is traced by the 975 A radiation field, to within
$\sim 10$\% (Go80,OW84), so that \CaII\ ionization rates should
increase with the 975 flux.  The increase of \NCaII\ with radiation
flux is predicted by the photoionization equilibrium of \CaII\ (see
the appendix). Predicted \NCaII\ values are plotted against the ISRF
for three different total H column densities, \NH=\NHI+\NHII, in
Fig. \ref{fig:caII}.  The observed increase of \NCaII\ with higher
fluxes is consistent with \CaII\ photoionization, and indicates that
higher electron densities and H II column densities are both
significant factors in the \CaII\ line strengths.

\section{Discussion and Conclusions} \label{sec:discussion}

The discussions in this paper are based on a search for evidence of
the S1 and S2 shells in local ISM data. The S1 and S2 shells are
assumed to be spherical and complete. Such simple assumptions
are justified only in the initial stage of probing
the ISM distribution associated with
a superbubble shell that has column densities too low
for \HI\ 21-cm measurements, and that has
evolved into a very low density region of space.
Several studies model the formation of the Local Bubble in terms
of the energy injected into the ISM by supernovae in the
Sco-Centaurus Association \citep{deGeus:1992,Frisch:1998LB,MaizApel:2001},
but the connection between the Loop I radio emission and very local ISM has
never been established.  The S1 and S2 shell models provide a basis for
testing this connection.

Early optical polarization data \citep{Tinbergen:1982} indicate that
the ISMF direction close to the Sun agrees with the S1 shell ISMF
direction once the uncertainties in Wolleben 's \Bphi\ angle (2007)
are included.  In principle \FeII\ can be used to trace the ISM
distribution, since is arises in both neutral and ionized
gas. Detailed comparisons between \FeII\ line strengths towards nearby
stars and the projected pathlength through the S1 and S2 shells
towards that star support, but do not prove, that the lines arise in
shells.  Both the \FeII\ and \CaII\ data indicate that the portions of
the S1 and S2 shells with \glong$>180^\circ$ will be more highly
ionized than in the opposite hemisphere.  The data are not sufficient
to distinguish an ionization gradient from an abundance gradient.  For
this reason, the shells are better traced using ions with first
ionization potentials less than 13.7 eV.  Heating by \lya\ radiation
accounts for $\sim 66$\% of the heating of the circumheliospheric ISM,
so shell regions exposed to the highest radiation flux should also be
warmer (an effect not explicitely included in the Ca equilibrium
discussion in the appendix).

These models of the S1 and S2 shells assume spherically symmetric
forms, which may be viable only for low density sightlines where
magnetic and thermal pressures are comparable.  Interstellar data are
compared mainly to the S1 shell, which has the most favorable geometry
for surrounding the Sun according to \citet{Wolleben:2007}.  The
Tinbergen data suggest a slow increase in polarizations with distance,
and the position angles towards the nose are consistent with the
optical polarization of more distant stars in Loop I
\citep{Frisch:2007cmb}.

The most distant shell regions, in the galactic center hemisphere,
have expanded into the high-extinction gas beyond $\sim 100$ pc that
is associated with the Sco-Cen Association (see e.g. Figs. 1,3 in
\citet{Vergelyetal:1998} or Fig. 2 in \citet{Frisch:2007grin}) and
show pronounced magnetic filaments (H98a) rather than a spherical
shell geometry.  Heiles points out that the synchrotron-emission
ridges of Loop I follow the distortion of the nearby global ISMF, as
traced by polarization data, and that Loop I is not a shell for the
high density regions.  The global ISMF within several hundered parsecs
is directed towards \glong$\sim 80^\circ - 88^\circ$.  The expansion
of the nearest portions of the S1 shell in a uniform field would yield
an ISMF close to the Sun directed upwards with respect to the galactic
plane.

\citet{HanayamaTomiska:2006snrsb} model the properties of a magnetic
superbubble $\sim 3$ Myrs old and for the strong ISMF case,
where magnetic pressure dominates thermal pressure by a factor of
$\sim 2$.  The superbubble cavity is elongated in the direction
parallel to the ISMF, where the shell is thinner.  The shell is
thicker and more extended in the radial direction, $\sim 90^\circ$
from the magnetic pole.  The region of strongest polarization for the
Tinbergen sample is $\sim 90^\circ$ from the magnetic pole.
Non-uniform shell expansion, or a 'wrinkled' shell, could explain the
anomalous sightlines towards HD 120315 and HD 80007 (see
Fig. \ref{fig:length} compared to Fig. \ref{fig:panels}), while the
tiny dense nearby Leo cloud \cite[][Lauroesch private
communication]{MeyerLauroeschHeiles:2006}, and the exceptionally
strong \CaII\ line towards Rasalhague (HD 159561, $\alpha$ Oph) may
indicate a region of merging shells.

The column densities for \FeII\ and \CaII\ are generally weaker for
sightlines with \glong$ < 180^\circ$, and stronger for stars with
\glong$ > 180^\circ$.  This effect may be either from the distribution
of ionized gas, or abundance variations for Fe and Ca.  The effect is
seen over small spatial scales of $\pm 10$ pc.  If the variation is
due to abundance differences, then the ISM close to the Sun would have
two different histories, although the flow velocities are similar.

The conclusions of this comparison between the S1 and S2 shells
with LISM markers can be briefly summarized:

\begin{itemize}

\item
The \citet{Wolleben:2007} description of the Loop I polarized radio
continuum in terms of two shells, S1 and S2, is viable and has
sufficient detail to be tested against observational data.  For
example, when the the S1 and S2 shell parameters are varied within the
allowed uncertainty range, the nearby cold gas filaments in Leo
\citep{Meyer:2007,Lauroesch:2007} are seen to be produced where the
two shells merge or collide (\S \ref{sec:shells}).

\item
The S1 shell magnetic field direction of \citet{Wolleben:2007}, with
\Bphi=--42\deeg, matchs the ISMF direction derived from older
polarization data \citep{Tinbergen:1982} of nearby stars near the
ecliptic plane and heliosphere nose, but offset by up to $\lambda \sim
+ 40^\circ$ from the heliosphere nose.  The ISMF direction implied by
the S1 shell and polarization position angles together is directed towards
\glong,\glat$=71^\circ,48^\circ$.

\item
For low column densities, log \NFeII$<12.5$ \cmtwo, the strength of
the \NFeII\ is better described by the star distance.  For higher
column densities, log \NFeII$>12.5$ \cmtwo, the strength of the
\NFeII\ is better described by the pathlength of the sightline through
the S1 and S2 shells. This result is based on a limited number of
stars ($<25$) and requires confirmation using a larger data set (\S
\ref{sec:feII}).

\item
The illumination of the S1 shell by the strong diffuse far ultraviolet interstellar
radiation field in Quadrants III and IV, \glong$>180^\circ$, explains
the higher column densities observed for \FeII\ and \CaII\ in these
galactic quadrants (\S \ref{sec:feII}, \S \ref{sec:caII}).
An appendix evaluates the ionization equilibrium of \CaII\
features spaced around the shell, and shows that the local radiation
field strength regulates the \CaII\ absorption line strengths.

\item
The ISMF direction at the heliosphere nose is within $10^\circ$ of
the angle of the great circle that is midway between
the hot and cold hemispheres of the CMB dipole moment,
and that also bifurcates the heliosphere nose.

\item
The S1/S2 shell model can be used to predict whether a star, or
exoplanet system for example, is embedded in a cloud or in the Local
Bubble plasma (\S \ref{sec:feII}).  The reverse is also true, that
measurements of astrosphere properties will help constrain the
distribution of ISM associated with the shells.

\item
This scenario describing the influence of the magnetic superbubble S1
and S2 shells on the local ISM, and as the origin of the interstellar
magnetic field at the Sun, is consistent with available data, but does
not yet prove the S1/S2 model.  Two kinds of data are required to
substantiate this picture: (1) Additional UV observations of tracers
of both neutral and ionized interstellar gas, e.g.  \FeII, \MgII, and
\MgI\ features.  (2) Measurements of nearby weak interstellar
polarizations at 0.01\% levels or better.

\end{itemize}

\acknowledgments The author would like to think NASA for research
funding, in the form of grants NAG5-13107 and NNG05GD36G to the
University of Chicago.
\appendix

\section{\CaII\ Equilibrium}

The \CaII\ column density can be determined from the assumption of
photoionization equilibrium between \CaII\ and \CaIII, for the
radiation flux level $F_\mathrm{i}$ at each star.  The highest fluxes
of diffuse far UV radiation are seen towards the third and fourth
galactic quadrants, \glong$>180^\circ$.  Some self-shielding of the
ISM in the two shells may occur, but the very low opacity of the Local
Bubble interior suggests that the $R^{-2}$ radial dependence of the
ISRF from the 25 brightest far UV stars is more germane for
understanding local ISM ionization, and the recombinant species \CaII\
that tracks the ionization.  For the temperature range considered
here, $3,000 - 15,000$ K, collisional ionization is insignificant
\citep{Pottasch:1972}.  The \CaII\ equilibrium depends on the \CaII\
photoionization rate $\Gamma_\mathrm{23}$, the recombination rate from
\CaIII\ to \CaII\ $\alpha_{32}$, the electron density \nel, the Ca
abundance $A_\mathrm{Ca}$, and the total hydrogen density $(\mathrm{
H^\circ ~ + ~ H^+ }) $:

\begin{equation} \label{eq:eq1}
\Gamma_{23} ~ \mathrm{Ca^+} ~ = ~ \alpha_{32} ~ \mathrm{n_e ~ Ca^{++}} 
\end{equation}

or

\begin{equation} \label{eq:eq1b}
\Gamma_{23} ~ \mathrm{Ca^+} ~ = ~ \alpha_{32} ~  \mathrm{n_e }~ (A \mathrm{_{Ca} ~ ( H^\circ + H^+) - Ca^+ ) }
\end{equation}

We define the ratio of the ionization and recombination rates at each star as
\begin{equation} \label{eq:eq2}
\Phi_\mathrm{i,j} ~ = ~ \Phi(F_\mathrm{i},T_\mathrm{j})  ~ = ~ \frac{\Gamma(F_\mathrm{i})}{\alpha(T_\mathrm{j})} =
~ \Phi_\mathrm{Mod26} * \frac{ F_\mathrm{i}}{F_{26}}  ~ (\frac{T_\mathrm{j}}{T_\mathrm{26}})^{0.8}
~=~ b ~ F_\mathrm{i} ~  T_\mathrm{j} ^ {0.8}
\end{equation}
for radiation flux $F_\mathrm{i}$, electron temperature
$T_\mathrm{j}$, $\alpha_{32} ~ \sim ~ T^{-0.8}$
\citep{ShullVanSteenberg:1985}, and after parameterizing $\Phi$ in
terms of the properties of the circumheliospheric ISM (CHISM) at the
solar location based on Model 26 in SF08.  For Model 26 in SF08,
\nel=0.0654 \cc\ (and \nHII=0.0554 \cc), \CaIII/\CaII=63.5,
$T_\mathrm{CHISM} ~ = ~ T_\mathrm{26}$=6320 K, giving
$\Phi_\mathrm{Mod26} = 4.15$ using eq. \ref{eq:eq1b}.  The ionization
edge of \CaII\ is at 1044 A.  For the ratio $F_\mathrm{i}/F_{26}$,
$F_\mathrm{i}$ is approximated by the total flux at each star $i$ from
the combined distance-corrected radiation fields of the 25 brightest
stars at 975 A \citep[based on fluxes as measured at the Sun, with no
opacity corrections, from][]{OpalWeller:1984}.  The normalization
factor $F_{26}$ is the 1044 A flux from \citet{Gondhalekaretal:1980}
as used for Model 26 in SF08, or $F_{26}= 80,000$ photons cm$^{-2}$
s$^{-1}$ A$^{-1}$.  With this scaling, $b = 4.74 \times 10^{-8}$ in
eq. \ref{eq:eq2}.  Calcium abundances appear to vary by a factor of
$\sim 40$ between cold and warm clouds \citep{Welty23:1999}. I use the
typical warm diffuse cloud calcium abundance of $A_\mathrm{Ca} =2.2
\times 10^{-8}$ calcium atoms per hydrogen atom.

\NCaII\ then becomes:

\begin{equation}\label{eq:eq3}
N(\mathrm{Ca^+}) = \frac{\mathrm{n_e} ~ \mathrm{A_{Ca}}~ N(\mathrm{H_{tot}})}{b ~ F_\mathrm{i} ~  T_\mathrm{j}^{0.8} + n_\mathrm{e} }
\end{equation}

\begin{figure}[h!]
\plottwo{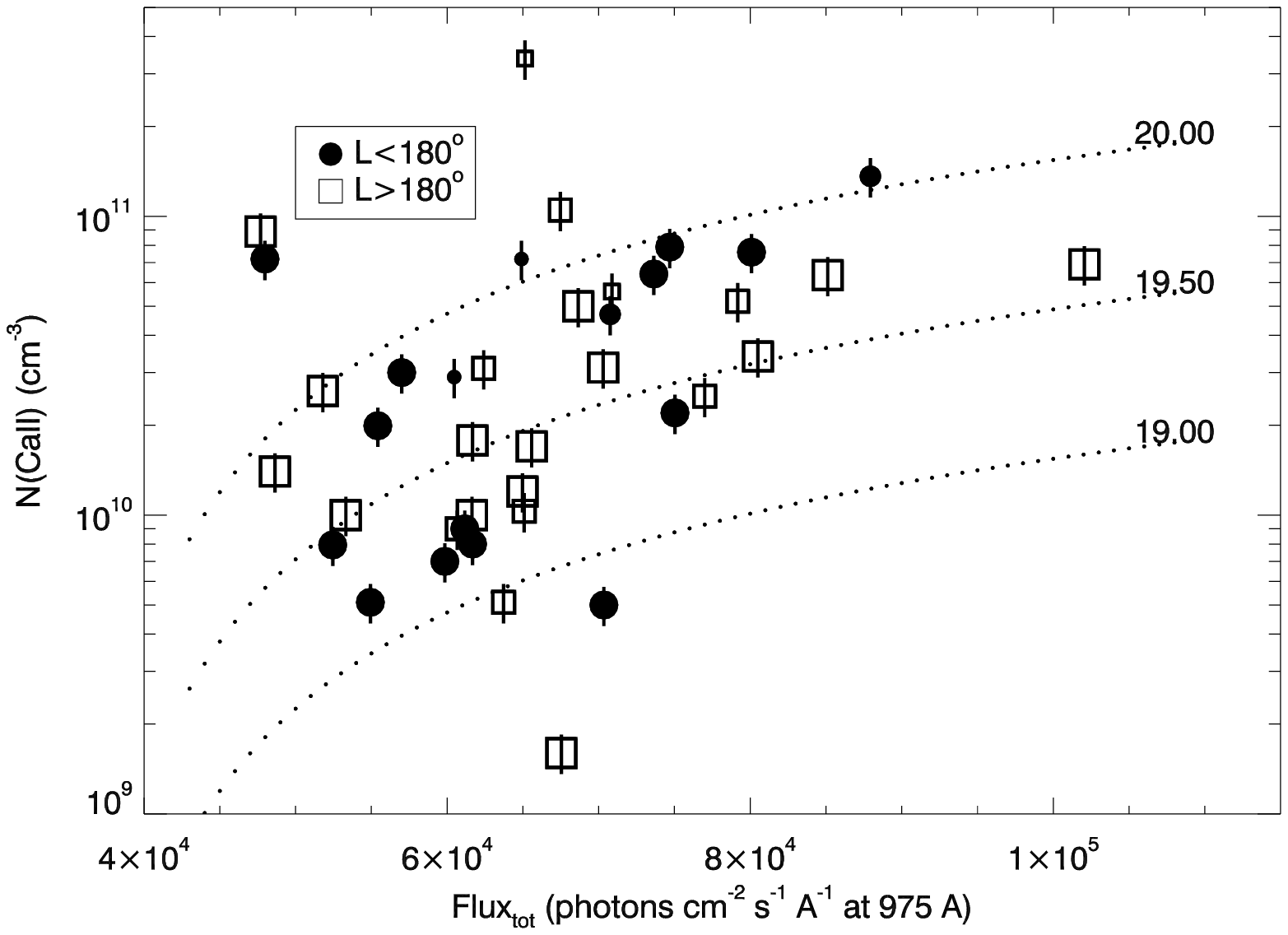}{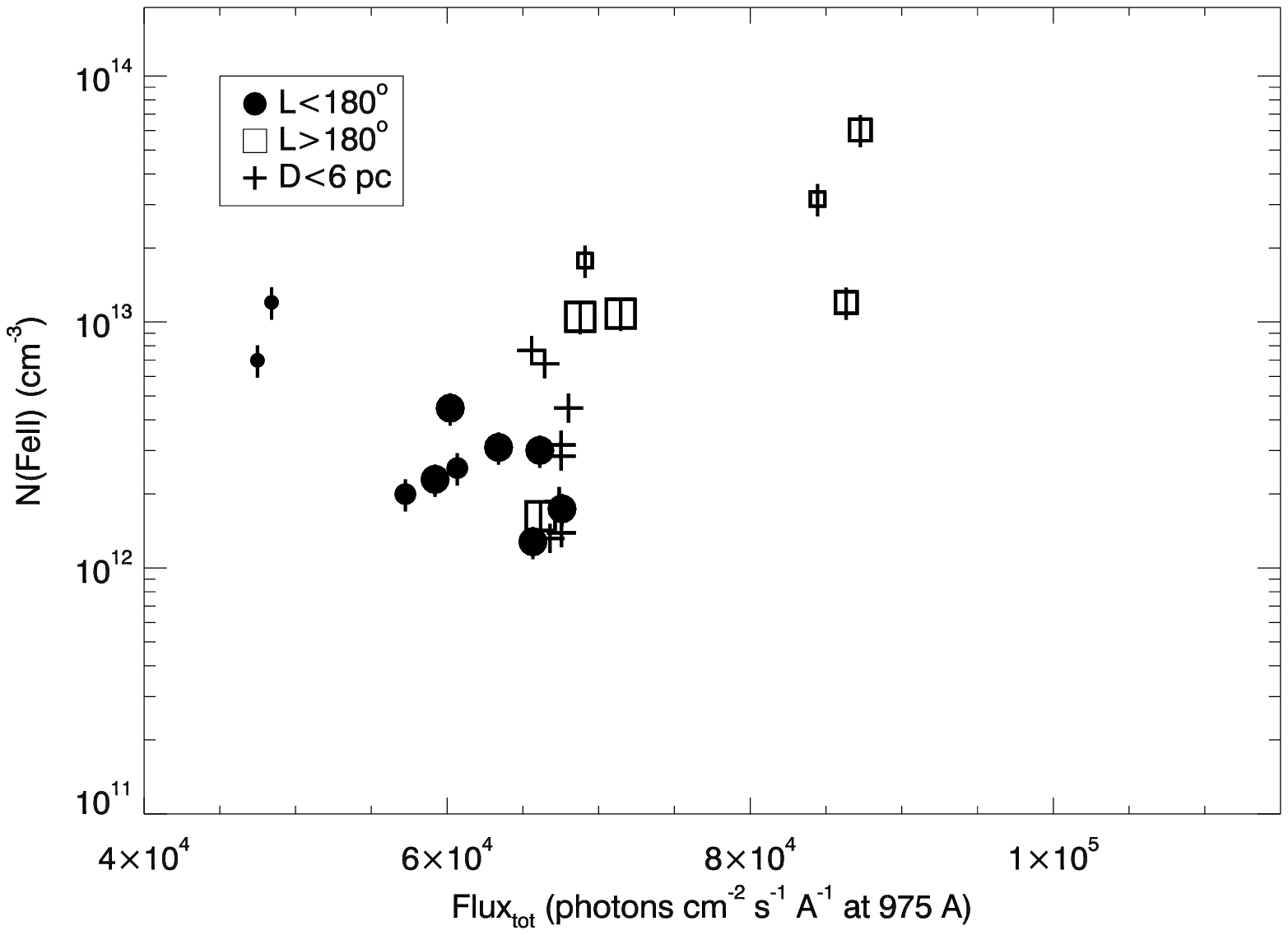}
\caption{The interstellar radiation flux at 975 A at the location of
each star is compared to \CaII\ column densities (left) and \FeII\
column densities (right).  The \CaII\ data are drawn from
\citep{FrischChoi:2008ip,FGW:2002}, and the ISRF data is from
\citet{OpalWeller:1984}.  Symbols indicate whether the galactic
longitude of the star is less than $180^\circ$ (filled), or greater
than $180^\circ$ (open).  The dotted curves show the predicted \CaII\
column densities for the labeled total hydrogen column densities
(\HI+\HII) for the assumption of photoionization equilibrium, and
assumptions for the Ca abundance ($2.2 \times 10^{-8}$ per hydrogen
atom) and temperature sensitivity to \nel\ (see appendix).  The two
stars with log \NCaII$\sim 10.8$ \cmtwo\ and low flux levels of $\sim
4.8 \times 10^4 ~
\mathrm{photons^{-2}~cm^{-2}~s^{-1}~A^{-1}}$ are between 50 and 56 pc
away.
\label{fig:caII} }
\end{figure}

The observed relation between \CaII\ column densities and the 975 A
flux seen in Fig. \ref{fig:caII} is compared to predicted values of
\NCaII\ determined from eq. \ref{eq:eq3}, for cloud temperatures in
the range of 2,500 and 15,000 K and electron densities in the range of
\nel=0.01--0.15 \cc\ and positive detections of \CaII\ for stars
within 55 pc \citep[using \CaII\ data from
][]{FGW:2002,FrischChoi:2008ip,WeltyCa:1996}\footnote{The \CaII\
column densities represent the sum of all components towards each star.
The stars are HD 358, HD 8538, HD 12311, HD 18978, HD 40183, HD 48915, HD
74956$^*$, HD 87901, HD 88955$^*$, HD 102124, HD 103287, HD 106591, HD
106625$^*$, HD 108767$^*$, HD 112413, HD 115892, HD 120315, HD
135742$^*$, HD 139006, HD 141003, HD 141378, HD 148857, HD 156164, HD
159561, HD 160613, HD 161868, HD 177724, HD 177756, HD 186882, HD
187642, HD 192696, HD 203280, HD 207098, HD 209952, HD 210418, HD
212061, HD 213558, HD 215789, HD 218045, and HD 222439.  The five
stars marked with an asterisk have the highest 975 A radiation fluxes
in this group.  }.  The missing parameter is the relation between the
cloud temperature and electron density, $T_\mathrm{j}$ in eq. A3, and
for that I use the somewhat arbitrary relation $T = 200 ~
n(\mathrm{e})^{-1.03}$ in order to account for the increased cooling
resulting from the collisional excitation of \CII\ and \OI\
fine-structure levels by electrons \citep{YorkKinahan:1979}.  This
assumption is required to fully specify the ionization equilibrium.
The lines labeled 20.00, 19.5, and 19.0 in Fig. \ref{fig:caII} show
the predicted \CaII\ column densities for these assumptions and log
\NH = log $N$(\HI+\HII) = 20.00, 19.5, and 19.0 \cmtwo.  If the gas is
clumpy, or conditions differ substantially from the CHISM gas, these
estimates will break down.  In the absence of a full 3D model of
opacity over several hundred parsescs, more detailed comparisons
between \NCaII\ and the ISRF require additional data on the cloud
temperature, H density, or ionization.  However this comparison
illustrates that reasonable assumptions for the parameters required to
calculate the ionization equilibrium of \CaII\ yield reasonable
predictions for the sensitivity of \NCaII\ to the far UV radiation
field.


\begin{deluxetable}{lccc}
\tablecolumns{4}
\tablecaption{Position Angles\label{tab:stars}}
\tablewidth{0pt}
\tablehead{
\colhead{Item} & \colhead{Galactic} & \colhead{Ecliptic} \\
\colhead{} & \colhead{Coords.} & \colhead{Coords.} \\
\colhead{} & \colhead{\PAgal (deg)} & \colhead{\PAecl (deg)} 
}
\startdata
Position angles for 3 upwind stars with $P>3\sigma$\tablenotemark{(A)} & $33 \pm 11$  &  $-26 \pm 11$  \\
Mean S1 shell $B$-field towards 3 upwind stars\tablenotemark{(B)}  & $39 \pm 6 $ & $-20 \pm 5 $    \\ 
\HI - \HeI\ offset at heliosphere nose (\glong,\glat$= 3.5^\circ ,15.2^\circ)$\tablenotemark{(C)}  & $26 \pm 20$  & $-35 \pm 20 $ \\
S1 shell $B-$field orientation at heliosphere nose \tablenotemark{(D)} & $46$   & $-15$     \\ 
Direction of CMB dipole midplane at (\glong,\glat)$ \sim (7.4^\circ, 11.6^\circ)$\tablenotemark{(E)} & $50 \pm 1$  & $-11 \pm 1$  \\
S1 shell $B-$field orientation  at (\glong,\glat)$ \sim (7.4^\circ, 11.6^\circ)^{(D)}$  & $43$ & $-18$     \\ 
\enddata
\tablenotetext{(A)}{These three stars are HD 161892, HD 177716, HD 181577, and the polarization
data are from \citet{Tinbergen:1982}.  The 
nominal uncertainties on the position angles are obtained by letting Q and U 
vary over $1\sigma$ measurment uncertainties of $\pm 0.007$\%.}
\tablenotetext{(B)}{The parameters for the S1 shell are given in \citet{Wolleben:2007}.}
\tablenotetext{(C)}{The \HI\ inflow direction is given in \citet{Lallementetal:2005}.
The \HeI\ B1950 inflow direction is given in \citet{Witte:2004}, and must be
corrected to J2000 coordinates (\S \ref{sec:tinb}).}
\tablenotetext{(D)}{This angle is calculated for \Btheta$=71^\circ$ and \Bphi$ \sim -42^\circ$,
and is within the $\pm 42^\circ$ uncertainties on the S1 shell ISMF direction
\citep[][after combining the $\pm 30^\circ$ uncertainties on each angle in quadrature.]{Wolleben:2007}}
\tablenotetext{(E)}{This location is the nearest point of the CMB dipole mid-plane to
the heliosphere nose direction, as defined by the interstellar \HeI\
flow through the heliosphere.  Uncertainties in the \HeI\ flow direction that
defines the heliosphere nose are also included in these comparison
uncertainties.  The CMB dipole directions are given in \citet{BennettBandayetal:1996}.
}
\end{deluxetable}
\clearpage

\clearpage

\clearpage

\clearpage

\clearpage

\clearpage

\end{document}